\documentclass[apj]{emulateapj}
\usepackage{times}
\usepackage[normalem]{ulem}
\usepackage{epsfig}
\usepackage{natbib}
\usepackage{amsfonts}
\usepackage{amsmath}
\usepackage{multirow}
\usepackage{enumerate}
\usepackage{verbatim}
\usepackage[usenames]{color}
\usepackage[plainpages=false, colorlinks=true, anchorcolor=blue, linkcolor=blue, citecolor=blue, bookmarks=false]{hyperref}
\newcommand{\be}{\begin{eqnarray}}
\newcommand{\ee}{\end{eqnarray}}

\newcommand{\lp}{\left(}
\newcommand{\rp}{\right)}
\newcommand{\lb}{\left[}
\newcommand{\rb}{\right]}

\newcommand{\slugcom}{Submitted for publication in The Astrophysical Journal}
\slugcomment{\slugcom}

\begin{document}

\normalsize


\title{The Fate of Neutron Star Binary Mergers}

\author{Anthony L. Piro\altaffilmark{1}}
\author{Bruno Giacomazzo\altaffilmark{2,3}}
\author{Rosalba Perna\altaffilmark{4}}

\altaffiltext{1}{Carnegie Observatories, 813 Santa Barbara Street, Pasadena, CA 91101, USA; piro@carnegiescience.edu}
\altaffiltext{2}{Physics Department, University of Trento, via Sommarive 14, I-38123 Trento, Italy}
\altaffiltext{3}{INFN-TIFPA, Trento Institute for Fundamental Physics and Applications, via Sommarive 14, I-38123 Trento, Italy}
\altaffiltext{4}{Department of Physics and Astronomy, Stony Brook University, Stony Brook, NY 11794, USA}

\begin{abstract}
Following merger, a neutron star (NS) binary can produce roughly one of three different outcomes: (1) a stable NS, (2) a black hole (BH), or (3) a supra-massive, rotationally-supported NS, which then collapses to a BH following angular momentum losses. Which of these fates occur and in what proportion has important implications for the electromagnetic transient associated with the mergers and the expected gravitational wave signatures, which in turn depend on the high density equation of state (EOS). Here we combine relativistic calculations of NS masses using realistic EOSs with Monte Carlo population synthesis based on the mass distribution of NS binaries in our Galaxy to predict the distribution of fates expected. For many EOSs, a significant fraction of the remnants are NSs or supra-massive NSs. This lends support to scenarios where a quickly spinning, highly magnetized NS may be powering an electromagnetic transient. This also indicates that it will be important for future gravitational wave (GW) observatories to focus on high frequencies to study the post merger GW emission. Even in cases where individual GW events are too low in signal to noise to study the post merger signature in detail, the statistics of how many mergers produce NSs versus BHs can be compared with our work to constrain the EOS. To match short gamma-ray burst (SGRB)  X-ray afterglow statistics, we find that the stiffest EOSs are ruled out.  Furthermore, many popular EOSs require $\sim60-70\%$ of SGRBs to be from NS-BH mergers rather than just binary NSs.
\end{abstract}


\section{Introduction}
\label{sec:intro}

The collection of binary neutron stars (NSs) in our Galaxy demonstrate
that these binaries are sufficiently compact to merge on
astrophysically-relevant timescales \citep[see the summary in][and
  references therein]{Martinez15}. Their mergers are among the most
powerful sources of gravitational waves (GWs) in the Universe, which are expected to
be detected and studied in the coming years by Advanced LIGO
\citep{LIGO}, Advanced Virgo \citep{VIRGO}, and KAGRA
\citep{KAGRA}. Binary NS mergers are also one of the best candidates for
short gamma-ray bursts \citep[SGRBs,][]{Nakar07}, which is supported by their presence
in old stellar environments, non-detection of an associated SN, and
large offsets from their host galaxies \citep{Berger14}. The two main
scenarios for generating the central engine are then the formation of
a strongly magnetized torus around a spinning black hole
\citep[e.g.,][]{Popham99,Lee07,Rezzolla11,Giacomazzo13a, Ruiz16} or the formation of
a long-lived magnetar \citep[e.g.,][]{Usov92,Zhang01,Metzger11,Giacomazzo13,Rowlinson13}. In
addition, post merger, binary NSs have been suggested for a variety of
electromagnetic transients that have not been conclusively
confirmed. This includes the outflow of neutron-rich material
producing a ``kilonova'' (\citealp{Metzger16}, and references therein,
although see \citealp{Berger13}; \citealp{Tanvir13}), the spin down of
a remnant magnetar producing high energy and later radio emission
\citep{Nakar11,Piro13,Metzger14}, and fast radio bursts (FRBs) from collapsing
supra-massive NSs \citep{Falcke14}. In still
other studies, it has been postulated that post-merger magnetars may
be an important source of ultrahigh-energy cosmic rays
\citep[UHECRs,][]{Piro16b}.

  \begin{deluxetable*}{lccccc}
  \tablecolumns{6} \tablewidth{460pt}
 \tablecaption{Equation of State and Merger Outcome Summary}
   \tablehead{ Equation of state & H4 & APR4 & GM1 & MS1 & SHT}
  \startdata
   Maximum $M_g$ (non-rotating) & 2.01 & 2.16 & 2.39 & 2.75 & 2.77 \\
   Maximum $M_b$ (non-rotating) & 2.30 & 2.61 & 2.83 & 3.30 & 3.33 \\
   Maximum $M_g$ (mass-shedding limit) & 2.38 & 2.58 & 2.87 & 3.29 & 3.33 \\
   Maximum $M_b$ (mass-shedding limit) & 2.70 & 3.07 & 3.36 & 3.89 & 3.94 \\
   Spin period (ms) & 0.74 & 0.53 & 0.68 & 0.70 & 0.72 \\
   $\beta=T/|W|$ & 0.108 & 0.137 & 0.130 & 0.134 & 0.137 \\
   Angular momentum ($GM_\odot^2c^{-1}$) & 3.60 & 4.70 & 5.68 & 7.60 & 7.84 \\
   NS remnants (\%) & 0.1 & 6.4 & 44.2 & 99.5 & 99.6 \\
   Supra-massive NS remnants (\%) & 23.2 & 74.3 & 55.5 & 0.5 & 0.4 \\
   BH remnants (\%) & 76.7 & 19.3 & 0.3 & 0 & 0
    \enddata
   \tablecomments{All masses are in units of $M_\odot$, and the spin periods, $\beta$, and angular momentum are all at the maximum mass, mass-shedding limit.}
\label{table}
\end{deluxetable*}

In all of these scenarios, a critical question is what is the resulting remnant left over from the binary NS merger. Roughly speaking, there are three main possibilities \citep[see][and references therein]{Faber12}. First, the two NSs may have sufficiently low masses to merge and just produce a stable NS. Second, the two merging NSs may have a combined mass above the maximum possible mass for NSs and collapse directly to form a black hole (BH). Finally, there is an in-between case where because of the large amount of angular momentum present in the binary, the merged remnant has sufficient spin to initially prevent collapse (the supra-massive NS case). Then, only once the remnant NS losses angular momentum (for example, though gravitational wave emission, neutrino driven winds, or magnetic dipole spin down), it finally collapses to a BH\footnote{There is also the fourth potential case where a remnant NS may be prevented from collapsing for $\sim10-100\,{\rm ms}$ by thermal or differential rotation effects (the ``hypermassive NS'' case), but we lump this with the direct collapse to BH scenario for the purposes of this work.}. The proportion with which each of these outcomes occur depends on the initial mass distribution of NS binaries, for which the Galactic sample is our best estimate \citep{Ozel12,Kiziltan13}, and the equation of state (EOS) of matter at ultrahigh densities, which continues to be an outstanding unsolved problem in high energy astrophysics \citep{Lattimer01}. Eventually, by comparing the expected distribution of outcomes for various EOSs with observed samples of events that are thought to correspond to different outcomes, strict constraints should be made on the possible EOSs.

In the future, the best way to do this will be by following the gravitational waves of binary NSs as they coalesce and merge. The coalescence will provide an estimate of each of the NS masses and the post merger signature will probe the remnant object, either corresponding to BH ringdown if a BH was promptly made or the vibrations of a remnant NS if the EOS will allow this \citep{Bauswein12}. On the other hand, given the high frequency of the post merger signal, it may be difficult with the advanced era of gravitational wave detectors to study this in full detail. Nevertheless, these experiments should still be able to tell whether a BH or NS remnant formed, which holds important information.

In addition, there may currently be evidence of what the merger outcome is from the early X-ray afterglows of SGRBs. Many of these events show an ``X-ray plateau'' that appears to come from extended engine activity, which is then followed by a rapid drop that has been argued to be too steep to be explained by an external shock model \citep{Rowlinson10}. A popular hypothesis for what causes this is that a supra-massive NS with a magnetar-like field has been created, which then powers the X-ray plateau until angular momentum losses cause it to collapse to a BH \citep{Rowlinson13,Lu14,Lu15}. If this is the case, then the proportion of SGRBs with X-ray plateaus should be an important constraint on EOSs \citep{Gao16}.

Motivated by these issues, here we theoretically explore the distribution of NS binary remnant scenarios for a variety of EOSs. In Section~\ref{sec:models}, we present relativistic calculations of NS masses for realistic EOSs and also discuss the importance of the difference between gravitational and baryonic mass when deriving the final remnant masses. In Section~\ref{sec:popsynth}, we describe our treatment of the NS mergers, including how we incorporate spin and assess which merger remnants result in supra-massive NSs.  In  Section~\ref{sec:grbs}, we use Monte Carlo techniques to merge a population of binary NSs and summarize the proportion of different outcomes. In Section~\ref{sec:compare}, we compare our results to constraints from SGRBs and discuss the various implications of our work, including the impact on gravitational wave searches and electromagnetic counterparts. Finally we conclude in Section~\ref{sec:discussion} with a summary of our results.

\section{Neutron Star Models}
\label{sec:models}

We model the NS structures using the EOSs summarized in Table \ref{table}. These have been chosen to span a representative range of maximum NS masses, but all with maximum gravitational masses $\gtrsim2\,M_\odot$ given current observational limits \citep{Demorest10,Antoniadis13}. For each of these EOSs we solve for the hydrostatic structure including general relativity and compute a grid of NS models spanning a wide range of masses. We calculate the structures for both a non-rotating NS and a maximally rotating NS for use in treating the merger remnant (further discussed in Section~\ref{sec:popsynth}). For the latter case we assume solid-body rotation because magnetic breaking eliminates differential rotation on an Alfv\'{e}n timescale of $\sim10-100\,{\rm ms}$ \citep{Baumgarte00,Shapiro00}.  All models have been computed using the LORENE library and in particular its publicly available codes {\tt rotseq} and {\tt nrotstar}. The former can easily compute sequences of non-rotating NSs, while the latter has been used to compute equilibrium configurations at the mass-shedding limit. The different EOSs were imported in LORENE in a table format either produced by using a piecewise polytropic approximation (APR4, \citealp{Endrizzi16}; H4, \citealp{Kawamura16}; MS1, \citealp{Ciolfi17}), provided by the EOS authors (SHT, \citealp{Kastaun16}), or given by the Compose project (GM1\footnote{http://compose.obspm.fr/}, \citealp{Glendenning91,Douchin01}). This large grid of models gives us for each mass and EOS a relation between the gravitational mass
\be
	M_g = 4\pi \int_0^R e r^2dr,
\ee
where $e$ is the energy density and $R$ is the NS radius, and the baryonic mass \citep{Shapiro83}
\be
	M_b = 4\pi \int_0^R \lb 1 - \frac{2Gm(r)}{rc^2}\rb^{-1/2}\rho r^2dr,
\ee
where $m(r)$ is the enclosed mass at a given radius and $\rho$ is the rest-mass density (both equations refer to non-rotating models, but similar ones can be derived for rotating models). The important point is that in general $M_b\gtrsim M_g$ because of the gravitational redshift factor.

\begin{figure}
\epsscale{1.2}
\plotone{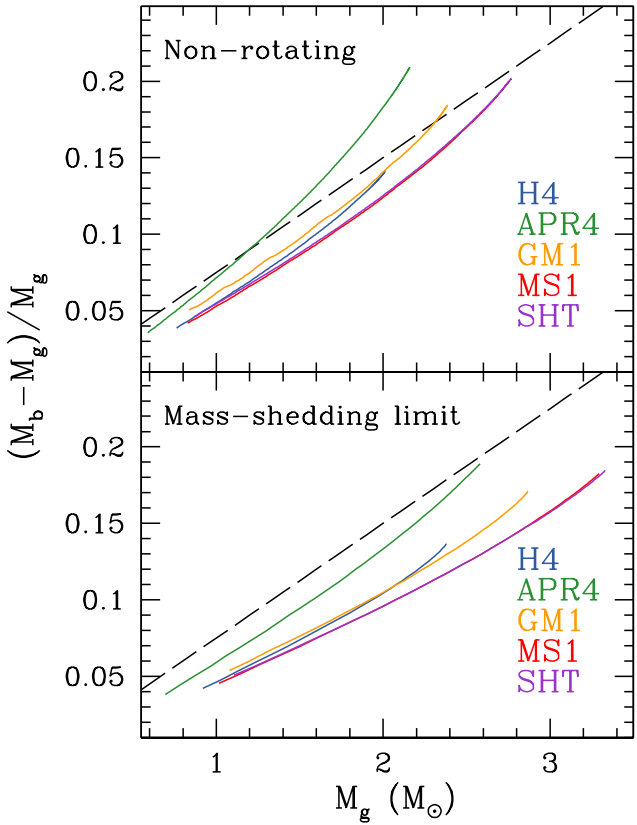}
\caption{The fractional difference between the baryonic mass $M_b$ and the gravitational mass $M_g$ as a function of $M_g$ for a range of EOSs. We compare both non-rotating (upper panel) and maximally rotating (lower panel) configurations, demonstrating that $M_b$ can be $\sim5-20\%$ higher than $M_g$. The black, dashed line shows Equation (\ref{eq:timmes}), an approximation commonly used in the literature \citep{Timmes96}.}
\label{fig:mg_mb}
\epsscale{1.0}
\end{figure}

The main results of these NS model calculations are summarized in Figure \ref{fig:mg_mb}. We plot the fractional difference between $M_b$ and $M_g$ for all of the models we have computed including both non-rotating and maximally rotating structures. This demonstrates that $M_b$ can be greater than $M_g$ by $\sim5-20\%$ with a larger difference for more massive NSs. In addition, not all EOSs have the same difference for a given $M_g$. This is because some EOSs have a smaller radius and thus are more relativistic (like APR4) than EOSs that give larger radii (like MS1). In addition, the difference is larger for non-rotating versus the rotating models, which will be crucial for understanding the final gravitational mass of merger remnants once it has spun down.

A commonly used approximation for the relation between $M_g$ and $M_b$ is \citep{Timmes96}
\be
	M_b = M_g + 0.075M_g^2,
	\label{eq:timmes}
\ee
where each of the masses are in solar units. This is also plotted in Figure \ref{fig:mg_mb} in comparison to our more detailed calculations. This demonstrates that the approximation has the correct trend, but generally overpredicts baryonic mass, especially at the mass-shedding limit. It also does not reflect the variation in this relation for different EOSs and at high spin.

\section{Treatment of Neutron Star Mergers}
\label{sec:popsynth}

Using this large grid of NS models we next explore the outcome when they merge. Consider NSs 1 and 2 with gravitational masses $M_{g,1}$ and $M_{g,2}$, respectively. During the inspiral, the NSs are not strongly affected by tidal coupling \citep{Bildsten92,Kochanek92}, and thus are not spun up appreciably. For this reason, their structures during this phase are well approximated by our non-spinning NS models.

Next, gravitational waves remove angular momentum from the binary until tidal disruption occurs for the lower mass NS. Assuming $M_{g,1}>M_{g,2}$, this occurs at a semi-major axis (\citealp{Kopal59}, adding the 10\% strong gravity correction from \citealp{Fishbone73})
\be
	a_t \approx 2.4 R_2\lp  1/q+1\rp^{1/3},
\ee
where $R_2$ is the radius of the less massive NS and $q=M_{g,2}/M_{g,1}$. The orbital angular momentum of the binary at this moment is
\be
	J_t &=& [G(M_{g,1}+M_{g,2})a_t]^{1/2}M_{g,2}
	\nonumber
	\\
	& \approx & 6.8 \lp 1/q+1 \rp^{1/3}\lp \frac{M_{g,2}}{1.4\,M_\odot}\rp^{3/2}
	\lp \frac{R_2}{10\,{\rm km}} \rp^{1/2} GM_\odot^2 c^{-1}
	\nonumber
	\\
	\label{eq:jt}
\ee
This amount of angular momentum must then be incorporated into the merger remnant unless significant high angular momentum material is flung away during the merger. In comparison, we summarize the angular momentum of our maximally spinning models in Table \ref{table}. Accounting for the fact that $(1/q+1)^{1/3}\gtrsim 1.25$ in Equation (\ref{eq:jt}), we see that the amount of angular momentum in the binary at the moment of tidal disruption always exceeds our maximally rotating models. Thus it is reasonable to assume that the merger remnant is near the mass shedding limit.

Given these estimates, our basic strategy is as follows. If we wish to merge NSs 1 and 2 with gravitational masses $M_{g,1}$ and $M_{g,2}$, respectively, we first convert these to their respective baryonic masses $M_{b,1}$ and $M_{b,2}$ using our grid of \emph{non-rotating} models. The total baryonic mass after the merger is then
\be
	M_{b,\rm tot} = M_{b,1} + M_{b,2} - M_{\rm lost},
\ee
where $M_{\rm lost}$ is the mass lost from the system during the merger. In general, this can be due to tidally stripped material, neutrino driven winds from an accretion disk, and neutrinos emitted by the hot remnant, but since the amount of mass lost is expected to be small \citep{Endrizzi16}, we use a fiducial value of $M_{\rm lost}=0.01\,M_\odot$ for this work. Then, using our \emph{maximally-rotating} models, we convert $M_{b,\rm tot}$ to the total resulting gravitational mass $M_{g,\rm tot}$. The motivation for using the rotating models when doing this conversion is the large angular momentum in the binary, as discussed above.

In cases where $M_{b,\rm tot}$ is greater than the maximum $M_b$ at the mass-shedding limit for a given EOS, we assume that remnant collapses to form a BH. If instead $M_{b,\rm tot}$ is less than  this maximum, but still above the maximum non-rotating $M_b$ ,we assume the remnant forms a supra-massive NS. This NS will only last until it loses sufficient angular momentum to collapse into a BH, potentially via magnetic dipole spin down or gravitational wave emission. Finally, if  $M_{b,\rm tot}$ is less than the maximum non-rotating $M_b$, then a stable NS is expected as the final product. In these cases with a stable NS, we perform one additional step of converting the $M_g$ in the mass-shedding limit to the (lower) non-spinning $M_g$, since this will be the final mass of the remnant once spin down has occurred.

The most likely angular momentum loss mechanisms are magnetic dipole spin down and GW emission. Thus the details of the spin down depend on exactly what magnetic fields are generated and what ellipticity is present, the latter of which can be caused by deforming the NS with internal toroidal magnetic fields. The associated timescales are discussed in detail in \citet{Lasky14}, who found that anywhere from $\ll1\,{\rm s}$ to $\sim10^3\,{\rm s}$ is possible depending on the exact parameters chosen. In some cases, it has even been argued that the spin down will proceed in steps to explain the plateau in emission in SGRBs (which we discuss further below), as initially GWs spin down the star on a relatively short timescale to a period of $\sim10\,{\rm ms}$, which is then followed by magnetic dipole emission over the next $\sim10^2-10^3\,{\rm s}$ \citep{Fan13}.

In principle another way to spin down the supra-massive NS remnants could be spin-related instabilities. These are typically controlled by the parameter $\beta=T/|W|$, where $T$ is the rotational energy and $W$ is the gravitational binding energy. For $\beta>0.27$, a dynamical bar-mode instability sets in, which leads to mass shedding and spin-down back to a stable state \citep[e.g.,][]{Shibata00}. For $0.27>\beta>0.14$, secular instabilities are possible, which can be triggered by viscosity and gravitational radiation reaction \citep{Chandrasekhar70,Friedman78,Lai01,Gaertig11}. In comparison, our maximally rotating models in Table \ref{table} always have \mbox{$\beta<0.14$,} so neither of these classes of instabilities are expected. Dynamical shear instabilities may operate for $\beta\gtrsim0.01$ \citep{Passamonti12,Passamonti13}, but this requires differential rotation which is not present in our models. We conclude that GWs and/or magnetic dipole emission are the main spindown mechanisms.

\begin{figure}
\epsscale{1.2}
\plotone{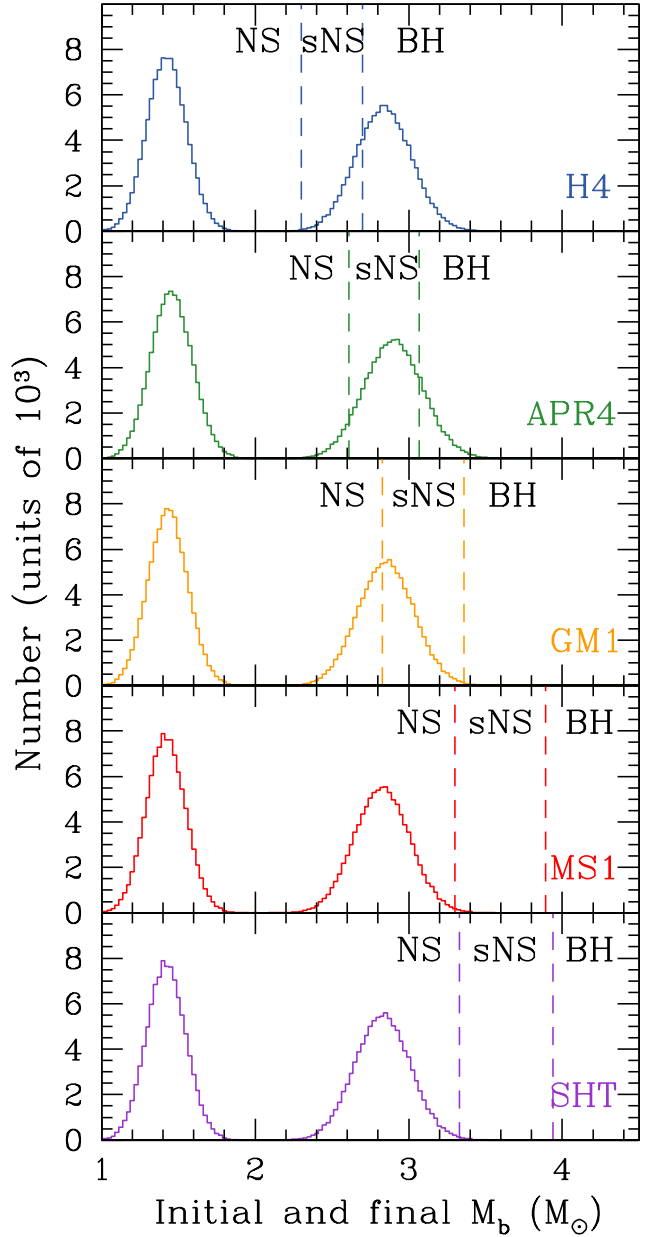}
\caption{Histograms of the initial (left distribution) and final (right distribution) baryonic masses of NSs after running $10^5$ binary models. The different colors correspond to different EOSs as noted in each panel. Vertical dashed lines indicate boundaries between where the final remnant will either be an NS, supra-massive NS (labeled as ``sNS''), or BH. The percentages of each of these outcomes is summarized in Table \ref{table}. }
\label{fig:histogram_bar}
\epsscale{1.0}
\end{figure}

\begin{figure}
\epsscale{1.2}
\plotone{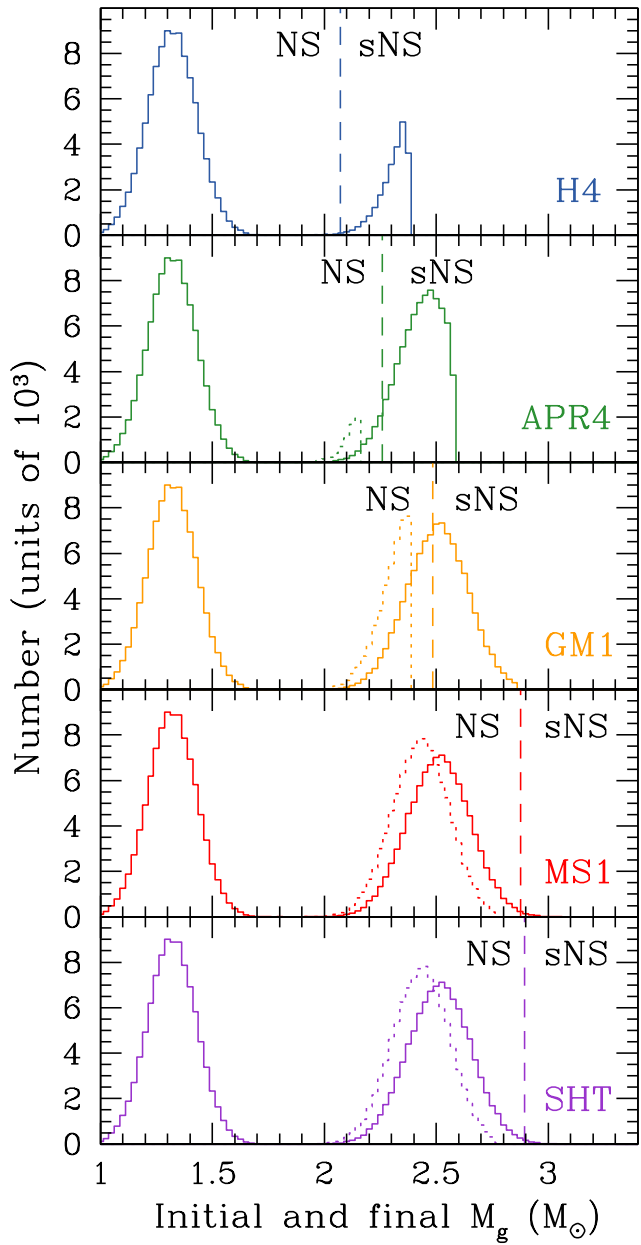}
\caption{Similar to Figure \ref{fig:histogram_bar}, but now showing the distribution of {\it gravitational} masses $M_g$. In this case, we cannot convert the remnant $M_b$ into $M_g$ if the final outcome is a BH, so these are left out of the final distribution. Furthermore, remnants that are stable NSs will have their $M_g$ decrease as they spin down. The final, distribution of these, in the limit of zero spin, is shown with the dotted lines.}
\label{fig:histogram_grav}
\epsscale{1.0}
\end{figure}

\section{Population Synthesis Results}
\label{sec:grbs}

Using the procedure outlined in Section \ref{sec:popsynth}, we next 
merge a population of NS binaries using each of the EOSs to find
the distribution of outcomes. In each case, we use Monte Carlo
techniques to produce $10^5$ binaries where the mass distribution
of the NSs is drawn from the galactic populations, which is
a Gaussian with a mean mass of $\mu=1.32\,M_\odot$ and
standard deviation of $\sigma=0.11\,M_\odot$ \citep{Kiziltan13}.
This is in contrast to other studies that attempt to perform the
population synthesis from a more first-principles perspective
including stellar models and binary interactions \citep[e.g.,][]{Fryer15}.

In Figure \ref{fig:histogram_bar}, we summarize the results of these
calculations for each of the EOSs. In each panel we plot the distribution
of initial (on the left) and final (on the right) baryonic masses
for the NS mergers. Even for the initial distributions of $M_b$ there
are subtle differences for each EOS because of the difference in the
conversion of initial $M_g$ values (which are all the same) to
$M_b$ (as summarized in Figure \ref{fig:mg_mb}).

The vertical dashed lines in Figure \ref{fig:histogram_bar} divide each
panel into three regions, showing the
criteria  that we use to assess the merger outcome. These
are either stable NSs, supra-massive NSs (labeled ``sNS''), or
BHs. These critical masses are given in Table \ref{table} as the
maximum $M_b$ (non-rotating) and maximum $M_b$ (mass-shedding
limit) rows. From the proportion of remnants in each of these three regions
we find the outcome percentages, summarized in the bottom three rows of Table \ref{table}.

These results demonstrate that even over this limited set of EOSs,
the distribution of outcomes can vary strongly. For H4, nearly all the
mergers eventually make BHs although many are first
supra-massive NSs. At the other extreme, MS1 and SHT make almost
all stable NSs. This indicates that measuring these distributions, for
example from SGRB features that are from one outcome or another,
strong constraints should be placed on the EOS.

\begin{figure}
\epsscale{1.2}
\plotone{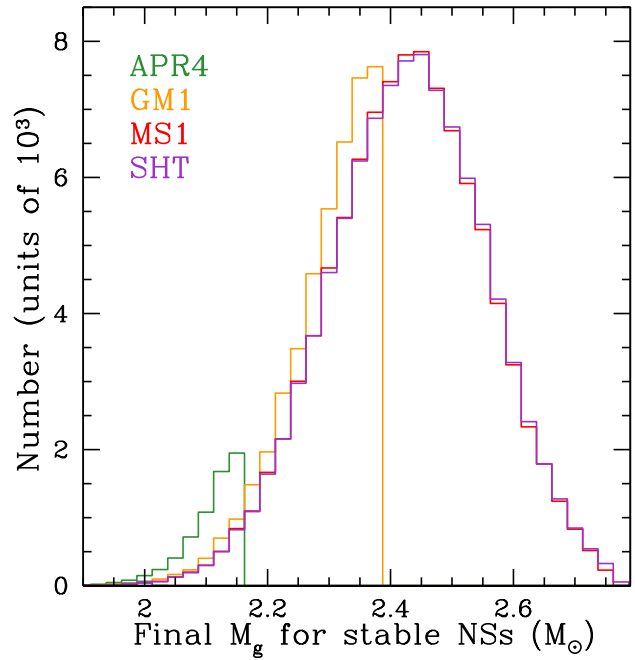}
\caption{The final $M_g$ distribution for the stable NSs following
spin down (the same as the dotted lines from Figure \ref{fig:histogram_grav}). H4 is not shown since it does not make appreciable
stable NSs.}
\label{fig:histogram_final}
\epsscale{1.0}
\end{figure}

Next, in Figure \ref{fig:histogram_grav} we summarize the
corresponding gravitational masses before (left distributions) and after (right
distribution) merger. In this case, we cannot calculate $M_g$ in the
cases where a BH is immediately formed, so these instances are
removed from the histograms on the right. This is why there is only
one dividing line between stable NSs and supra-massive NSs. This
line is given by taking the maximum, non-rotating $M_b$ from Table~\ref{table}
and converting it to $M_g$ in the mass-shedding, which is not given
in the table. Eventually the NSs on the left of the vertical line will spin down
to become stable NSs. As this occurs, their $M_g$ decreases because
they become more relativistic as their radius shrinks. Therefore the final
distribution of $M_g$ for the stable NSs are shown with dotted lines.
This is further highlighted in Figure \ref{fig:histogram_final} to focus
on these distributions.

\section{Comparisons and Implications}
\label{sec:compare}

The distribution of fates we find has a number of important connections to
SGRBs, other electromagnetic counterparts, gravitational wave signatures,
and ultrahigh-energy cosmic rays. We discuss each of these below with particular emphasis on
instances where these observations can be used to constrain the EOS.

\subsection{Short Gamma-ray Bursts}

As introduced in Section \ref{sec:intro}, several mixed pieces of evidence point to the
merger of two compact objects in a binary (NS-NS or NS-BH) as the
engine powering SGRBs. In the cases with a NS remnant,
it is often thought that it will have magnetar-strength magnetic field
($\sim10^{14}-10^{15}\,{\rm G}$),
either generated by an initially weak field amplified by shear instabilities at the merger interface
\citep{Price06,Zrake13} or by an $\alpha-\Omega$ dynamo in the subsequent neutrino cooling
phase \citep{Duncan92}. This can then potentially power additional emission via magnetic
dipole spin down.
Indeed, the early observations by the
{\em Swift} satellite of extended X-ray plateaus followed by either
power-law decays or rapid decays suggest that some of these mergers
leave behind either a stable NS, or a supra-massive NS, which powers this emission
until it collapses to a BH after slowing down.

Using such arguments, \citet{Gao16} analyzed 96 SGRBs observed with {\em Swift}
between January 2005 and October 2015, and found that
21 require a supra-massive NS, a fraction of 22\%. Our modeling would
argue that the EOS H4 matches best with this number. Nevertheless, given
the uncertainties in securely identifying whether a supra-massive NS is
needed and the limited number of events, GM1 might also be possible, and
indeed \citet{Gao16} conclude from their analysis that GM1 is favored.
Interestingly, H4 and GM1 give very different predictions for the fraction
of stable NSs and prompt BHs.

\begin{figure}
\epsscale{1.2}
\plotone{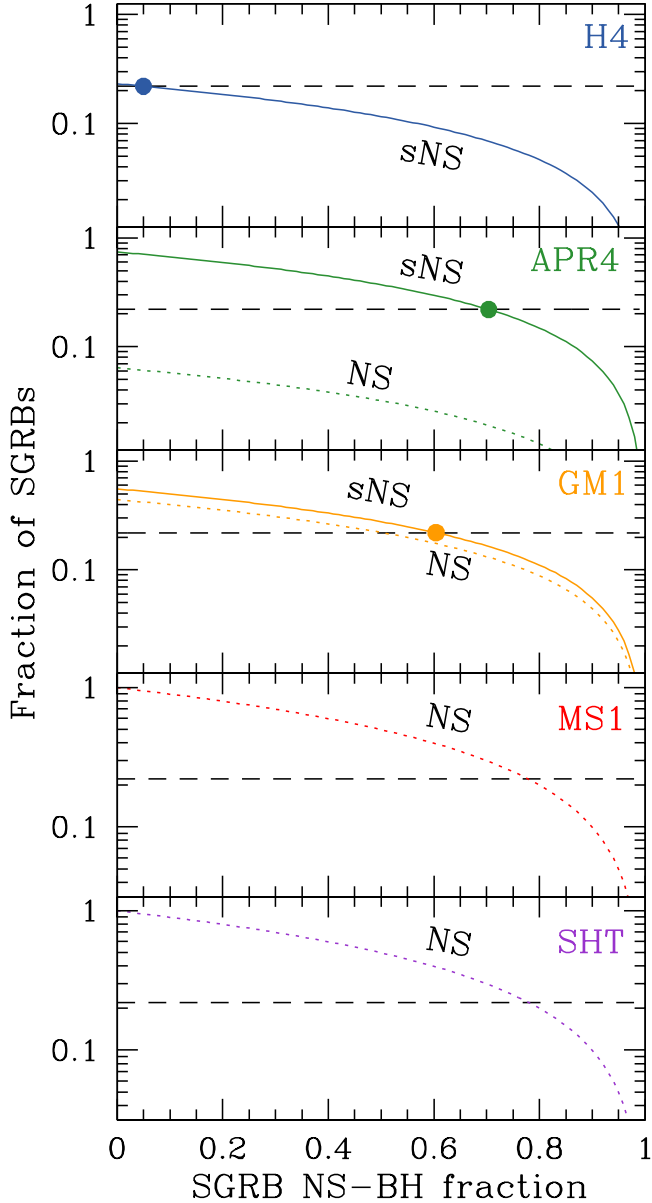}
\caption{The merger remnant fraction for SGRBs as a function of the fraction of
SGRBs that are from NS-BH mergers. For each EOS, the solid lines are the fraction of supra-massive
NSs and the dotted lines are the fraction of stable NSs. The black dashed line at $22\%$ is the fraction of
supra-massive NSs as argued by \citet{Gao16}. Filled circles show where we find each EOS can
match this fraction, demonstrating a significant amount of NS-BH mergers is required in
some cases.}
\label{fig:fraction}
\epsscale{1.0}
\end{figure}

However, even though we have so far only focused on NS-NS binaries,
it is possible that a fraction of SGRBs is due to NS-BH
mergers as well. Those cases would be added to (and confused with) the sample of
prompt collapse and dilute the inferred number of supra-massive NSs
made from NS-NS binaries. In Figure \ref{fig:fraction}, we plot the fraction of SGRBs
that result in NS and supra-massive NS remnants as a function of the
SGRBs that are due to NS-BH mergers. Quite simply, as the contribution from
NS-BH mergers increase, the others must go down. The filled circles denote where the other
EOS may match the SGRB constraints. For example, GM1 can get the observed 22\%
if $\sim$~60\% of SGRBs are from NS-BH mergers, and
APR4 could work if $\sim$~70\% of SGRBs are from NS-BH mergers.
On the other hand, there is no way to
produce appreciable supra-massive NSs with MS1 or SHT, so the stiffest EOSs
appear ruled out if the interpretation of these SGRB afterglows is correct.

Although the number of events is lower than in \citet{Gao16}, in the earlier work by \citet{Rowlinson13},
they also argue for a significant fraction of NS remnants from SGRBs. In this case, they conclude
that roughly $\sim64\%$ of SGRBs
form
some kind of NS and $\sim30-60\%$ of events form a supra-massive NS. Similar to what was concluded above,
the best fits are
APR4 and GM1, but only if again a significant fraction of $\sim60\%$ of SGRBs are from NS-BH mergers.

\subsection{Other Electromagnetic Transients}

Although SGRBs are seen and thought to be associated with NS-NS or
NS-BH mergers, there are a number of other electromagnetic transients
possibly associated with these events \citep[see the review by][]{Metzger12}.
Many of these emission mechanisms
are more isotropic than the relativistic jet of a SGRB, and thus promising
for detection, especially in the near future with surveys with especially
rapid cadences, such as the Zwicky Transient Facility \citep[ZTF,][]{Law09},
the All-sky Automated Survey for Supernovae \citep{Shappee14},
or the Large Synoptic Survey Telescope
(\citealp{lsst}, depending on the final cadence).

In particular, there are the scenarios that require an NS or supra-massive
NS remnant following the merger. Some examples include the post-merger
pulsar wind nebula that can produce an optical and X-ray signature on
short timescales of days \citep{Metzger14,Siegel16a,Siegel16b} and radio emission
on the timescale of months \citep{Piro13}. Additionally, radio emission may 
be possible from the small amount of material ejected by the merger ($\sim0.01\,M_\odot$)
that is accelerated by the energy input of a magnetar, which then interacts with
the nearby circumstellar medium \citep{Nakar11}. In any case, given that the SGRB comparison
favors the EOSs H4, APR4, or GM1, this would argue that these sorts of transients
are not uncommon and that future transient surveys can make important
constraints on whether these events are occurring.
The fraction of SGRBs with radio emission long after the main
burst could be combined with our results to constrain the EOS
\citep[e.g.,][]{Metzger14b,Fong16}, although one must be careful to
include that some SGRBs may be NS-BH mergers (see Figure \ref{fig:fraction}).
This could help decide which of H4, APR4, or GM1 is favored since for the combined fraction
of NSs and supra-massive NSs, they give very different results of
23.3\%, 80.7\%, and 99.7\%, respectively.

Another scenario to consider is whether these mergers make
FRBs, a recently discovered class of transients characterized by
millisecond bursts of radio emission
\citep{Lorimer07,Keane12,Thornton13,Ravi15}.
One idea for generating these is through the collapse of a
supra-massive NS, which expels its magnetic field as it becomes
a BH \citep{Falcke14}. Given our work here, it appears that
this scenario is quite plausible following a binary NS merger for
many EOSs. Nevertheless, it is probably difficult for
this to produce the majority of FRBs. Even the highest rate
of supra-massive NSs by APR4 will only produce roughly
$\sim1\%$ of FRBs \citep{Rane16,Nicholl17}. Furthermore,
the repeating FRB 121102 \citep{Spitler14,Spitler16,Scholz16}
cannot be reconciled with such a scenario where the FRB is
generated in a single catastrophic event.

\subsection{Impact on Gravitational Wave Searches}

It is expected with the new generation of gravitational wave detectors
that there will eventually be the detection of binary NS mergers.
From these, observations of relatively nearby events could help
constrain the EOS via the impact of tidal interactions during the
late stages of the merger
\citep[e.g.,][]{Flanagan08,Read09,Hinderer10,Baiotti10,Bernuzzi12}.
Since according to our analysis a significant fraction of NS mergers
will produce a long-lived NS remnant, we should
expect them to be accompanied by a long-duration post-merger GW signal
that will be important for EOS constraints as well.

When an NS or
supra-massive NS is generated, the frequency structure
holds detailed clues about the EOS \citep[see][and referencers therein]{Dall'Osso15,Clark16}
in the range of $1-4\,{\rm kHz}$. Due to decreased sensitivity
of the advanced gravitational wave detectors in this range, detection
of these features will either require especially nearby events, potential
upgrades to these detectors focused at high frequencies
\citep{LIGO,Miller15}, or the
Einstein Telescope \citep{Punturo10,Hild11}. Alternatively,
our work shows that the actual fraction of outcomes could also be constraining
for the EOS. This would be useful in the case that there is a large number of events,
but their individual post-merger frequency structures are
too low signal to noise to constrain the EOS from any given single
event. This is likely to be expected because the BH ringdown for
this mass range is much too high for the current generation of GW detectors,
while the signature of an NS remnant might still be observed.
By simply identifying which events formed
NSs (and/or supra-massive NSs) rather than BHs, the proportion
of outcomes could rule out some EOSs using our results as summarized
in Table \ref{table}.

\subsection{Ultrahigh-energy Cosmic Rays}

\citet{Piro16b} argue that rapidly spinning magnetars born from NS mergers
are a promising site for the generation of UHECRs
(particles with individual energies exceeding $10^{18}\,{\rm eV}$, see
reviews by
\citealp{Kotera11,Letessier11}). This
is because the short-period spin and strong magnetic fields
of an NS remnant
are able to accelerate particles up to appropriate energies, and the
composition of material on and around the NS may naturally
explain recent inferences of heavy elements in UHECRs.
The low ejecta mass in this scenario (in contrast to UHECRs
originating from the centers of supernovae \citealp[e.g.,][]{Arons03,Fang12,Kotera15})
also helps the high-energy neutrino background associated with this
scenario to be below current constraints from the IceCube
Observatory \citep{IceCube,Aartsen13}.

A critical uncertainty in this work is whether the fate of the
binary NS mergers can explain the detection rate of UHECRs. For binary NS mergers,
\citet{Kim05} estimates a volumetric rate of $\sim4\times10^{-8}-2\times10^{-6}\,{\rm Mpc^{-3}\,yr^{-1}}$,
which \citet{Piro16b} argue could explain the UHECR rate of
$\approx10^{43.5}-10^{44}\,{\rm erg\,Mpc^{-3}\,yr^{-1}}$
\citep{Waxman95,Berezinsky06,Katz09,Murase09} if
$\sim10-100\%$ of the mergers results in NSs or supra-massive NSs.
In fact, this is accommodated by many of the models we consider, namely H4, APR4, and GM1.
Furthermore, the high spin of these remnants is especially important for
accelerating particles to the required energies. Paradoxically, the magnetic field need not
be at the highest strengths possible. This is because a somewhat lower field
 of $\sim10^{13}-10^{14}\,{\rm G}$ (as opposed to $\gtrsim10^{15}\,{\rm G}$)
will delay the injection of UHECRs until the surrounding debris has
inflated sufficiently and is less likely to impede the particles from leaving
the remnant. Ultimately, the detection of binary NS mergers as GW sources will 
constrain this rate, so that the viability of this model for UHECRs can
be better tested.

\section{Discussion and Conclusions}
\label{sec:discussion}

We have investigated the distribution of outcomes for NS binary mergers
by combining relativistic calculations of spinning and non-spinning NS
structures for realistic EOSs with the Galactic binary NS mass distribution
using Monte Carlo methods. This demonstrated that the fate of NS mergers
can vary greatly depending on the EOS, from primarily supra-massive
NSs and prompt BHs for EOSs like H4 and primarily stable NSs for EOSs
like MS1 and SHT. In still other cases, a fairly even distribution of outcomes
is possible.

Perhaps most striking from our results is the non-negligible fraction of NS or
supra-massive NS remnants expected across all EOSs. At the low end, H4
predicts 23\% supra-massive NSs, but when taking NSs and supra-massive
NSs together, the other four EOSs we consider predict a fraction $\sim80-100\%$.
This means that it will be especially important for future GW observatories to
be able to probe the high frequency GW emission in the range of $1-4\,{\rm kHz}$
to study the frequency structure of the post-merger NS or supra-massive
NS remnant. Even if particular events are too low in signal to noise to study this
frequency spectrum in detail, \emph{if the statistics of NS versus BH remnants can be
measured, these proportions can be compared to our results to constrain the EOS.}

This means that observations that are sensitive to the ratio of these various
outcomes can be useful tools for constraining the EOS. In particular, we have
discussed the plateau signature in SGRBs X-ray afterglows. If they require $\approx22\%$ of
events to result in a supra-massive NS \citep[as argued by][]{Gao16}, then 
potentially H4, APR4, or GM1 can explain this.
Interestingly, if one of these latter two EOSs
end up being correct, then \emph{a significant fraction of $\sim60-70\%$ of SGRBs must be from NS-BH
mergers.} The inference of such a large number of NS-BH mergers is exciting given that there has
never been a direct observation of such a system. This has a number
of important implications. First, the rate of SGRBs \citep[corrected for beaming,
as estimated in][]{Fong12} could be greater than the rate of just binary
NS mergers \citep[e.g.,][]{Kalogera04}, and we may expect NS-BH mergers to be a significant
GW source in the near future.
Next, NS-BH mergers cannot have BH masses that are too large, otherwise
there may not be sufficient material for an accretion disk to power
the prompt SGRB emission \citep[$\lesssim8-10\,M_\odot$,][and references therein]{Shibata11,Foucart12,Rosswog15}.
This affects the production of {\it r}-process material from the mergers
and the rate of SGRBs need to produce the observed
abundances \citep[for reviews, see][]{Arnould07,Qian07,Sneden08,Thielemann11}.
The luminosity of
the radioactive electromagnetic transients from compact mergers \citep[e.g., kilonovae,][]{Metzger16}
would also be affected,
which will be studied by the upcoming wide field  surveys that are
sensitive to short timescale transients \citep[e.g., ZTF,][]{Law09},

\acknowledgments
We thank the OCIW Distinguished Visitor Program at the Carnegie Observatories, which funded R.P.'s visit where the initial ideas for this work were developed. R.P. acknowledges partial support from the NSF award AST-1616157. B.G. acknowledges partial support from ``NewCompStar'' COST Action MP1304 and from MIUR FIR grant No.~RBFR13QJYF. Some of the computations were performed on the MIES cluster of the Carnegie Observatories, which was made possible by a grant from the Ahmanson Foundation.

\bibliographystyle{apj}

\begin{thebibliography}{110}
\expandafter\ifx\csname natexlab\endcsname\relax\def\natexlab#1{#1}\fi

\bibitem[{{Aartsen} {et~al.}(2013){Aartsen}, {Abbasi}, {Abdou}, {Ackermann},
  {Adams}, {Aguilar}, {Ahlers}, {Altmann}, {Auffenberg}, {Bai}, \&
  et~al.}]{Aartsen13}
{Aartsen}, M.~G., {Abbasi}, R., {Abdou}, Y., {et~al.} 2013, \apj, 779, 132

\bibitem[{{Acernese} {et~al.}(2015){Acernese}, {Agathos}, {Agatsuma}, {Aisa},
  {Allemandou}, {Allocca}, {Amarni}, {Astone}, {Balestri}, {Ballardin}, \&
  et~al.}]{VIRGO}
{Acernese}, F., {Agathos}, M., {Agatsuma}, K., {et~al.} 2015, Classical and
  Quantum Gravity, 32, 024001

\bibitem[{{Antoniadis} {et~al.}(2013){Antoniadis}, {Freire}, {Wex}, {Tauris},
  {Lynch}, {van Kerkwijk}, {Kramer}, {Bassa}, {Dhillon}, {Driebe}, {Hessels},
  {Kaspi}, {Kondratiev}, {Langer}, {Marsh}, {McLaughlin}, {Pennucci}, {Ransom},
  {Stairs}, {van Leeuwen}, {Verbiest}, \& {Whelan}}]{Antoniadis13}
{Antoniadis}, J., {Freire}, P.~C.~C., {Wex}, N., {et~al.} 2013, Science, 340,
  448

\bibitem[{{Arnould} {et~al.}(2007){Arnould}, {Goriely}, \&
  {Takahashi}}]{Arnould07}
{Arnould}, M., {Goriely}, S., \& {Takahashi}, K. 2007, \physrep, 450, 97

\bibitem[{{Arons}(2003)}]{Arons03}
{Arons}, J. 2003, \apj, 589, 871

\bibitem[{{Aso} {et~al.}(2013){Aso}, {Michimura}, {Somiya}, {Ando}, {Miyakawa},
  {Sekiguchi}, {Tatsumi}, \& {Yamamoto}}]{KAGRA}
{Aso}, Y., {Michimura}, Y., {Somiya}, K., {et~al.} 2013, \prd, 88, 043007

\bibitem[{{Baiotti} {et~al.}(2010){Baiotti}, {Damour}, {Giacomazzo}, {Nagar},
  \& {Rezzolla}}]{Baiotti10}
{Baiotti}, L., {Damour}, T., {Giacomazzo}, B., {Nagar}, A., \& {Rezzolla}, L.
  2010, Physical Review Letters, 105, 261101

\bibitem[{{Baumgarte} {et~al.}(2000){Baumgarte}, {Shapiro}, \&
  {Shibata}}]{Baumgarte00}
{Baumgarte}, T.~W., {Shapiro}, S.~L., \& {Shibata}, M. 2000, \apjl, 528, L29

\bibitem[{{Bauswein} {et~al.}(2012){Bauswein}, {Janka}, {Hebeler}, \&
  {Schwenk}}]{Bauswein12}
{Bauswein}, A., {Janka}, H.-T., {Hebeler}, K., \& {Schwenk}, A. 2012, \prd, 86,
  063001

\bibitem[{{Berezinsky} {et~al.}(2006){Berezinsky}, {Gazizov}, \&
  {Grigorieva}}]{Berezinsky06}
{Berezinsky}, V., {Gazizov}, A., \& {Grigorieva}, S. 2006, \prd, 74, 043005

\bibitem[{{Berger}(2014)}]{Berger14}
{Berger}, E. 2014, \araa, 52, 43

\bibitem[{{Berger} {et~al.}(2013){Berger}, {Fong}, \& {Chornock}}]{Berger13}
{Berger}, E., {Fong}, W., \& {Chornock}, R. 2013, \apjl, 774, L23

\bibitem[{{Bernuzzi} {et~al.}(2012){Bernuzzi}, {Nagar}, {Thierfelder}, \&
  {Br{\"u}gmann}}]{Bernuzzi12}
{Bernuzzi}, S., {Nagar}, A., {Thierfelder}, M., \& {Br{\"u}gmann}, B. 2012,
  \prd, 86, 044030

\bibitem[{{Bildsten} \& {Cutler}(1992)}]{Bildsten92}
{Bildsten}, L., \& {Cutler}, C. 1992, \apj, 400, 175

\bibitem[{{Chandrasekhar}(1970)}]{Chandrasekhar70}
{Chandrasekhar}, S. 1970, \apj, 161, 561

\bibitem[{{Ciolfi} {et~al.}(2017){Ciolfi}, {Kastaun}, {Giacomazzo}, {Endrizzi},
  {Siegel}, \& {Perna}}]{Ciolfi17}
{Ciolfi}, R., {Kastaun}, W., {Giacomazzo}, B., {et~al.} 2017, \prd, 95, 063016

\bibitem[{{Clark} {et~al.}(2016){Clark}, {Bauswein}, {Stergioulas}, \&
  {Shoemaker}}]{Clark16}
{Clark}, J.~A., {Bauswein}, A., {Stergioulas}, N., \& {Shoemaker}, D. 2016,
  Classical and Quantum Gravity, 33, 085003

\bibitem[{{Dall'Osso} {et~al.}(2015){Dall'Osso}, {Giacomazzo}, {Perna}, \&
  {Stella}}]{Dall'Osso15}
{Dall'Osso}, S., {Giacomazzo}, B., {Perna}, R., \& {Stella}, L. 2015, \apj,
  798, 25

\bibitem[{{Demorest} {et~al.}(2010){Demorest}, {Pennucci}, {Ransom}, {Roberts},
  \& {Hessels}}]{Demorest10}
{Demorest}, P.~B., {Pennucci}, T., {Ransom}, S.~M., {Roberts}, M.~S.~E., \&
  {Hessels}, J.~W.~T. 2010, \nat, 467, 1081

\bibitem[{{Douchin} \& {Haensel}(2001)}]{Douchin01}
{Douchin}, F., \& {Haensel}, P. 2001, \aap, 380, 151

\bibitem[{{Duncan} \& {Thompson}(1992)}]{Duncan92}
{Duncan}, R.~C., \& {Thompson}, C. 1992, \apjl, 392, L9

\bibitem[{{Endrizzi} {et~al.}(2016){Endrizzi}, {Ciolfi}, {Giacomazzo},
  {Kastaun}, \& {Kawamura}}]{Endrizzi16}
{Endrizzi}, A., {Ciolfi}, R., {Giacomazzo}, B., {Kastaun}, W., \& {Kawamura},
  T. 2016, Classical and Quantum Gravity, 33, 164001

\bibitem[{{Faber} \& {Rasio}(2012)}]{Faber12}
{Faber}, J.~A., \& {Rasio}, F.~A. 2012, Living Reviews in Relativity, 15, 8

\bibitem[{{Falcke} \& {Rezzolla}(2014)}]{Falcke14}
{Falcke}, H., \& {Rezzolla}, L. 2014, \aap, 562, A137

\bibitem[{{Fan} {et~al.}(2013){Fan}, {Wu}, \& {Wei}}]{Fan13}
{Fan}, Y.-Z., {Wu}, X.-F., \& {Wei}, D.-M. 2013, \prd, 88, 067304

\bibitem[{{Fang} {et~al.}(2012){Fang}, {Kotera}, \& {Olinto}}]{Fang12}
{Fang}, K., {Kotera}, K., \& {Olinto}, A.~V. 2012, \apj, 750, 118

\bibitem[{{Fishbone}(1973)}]{Fishbone73}
{Fishbone}, L.~G. 1973, \apj, 185, 43

\bibitem[{{Flanagan} \& {Hinderer}(2008)}]{Flanagan08}
{Flanagan}, {\'E}.~{\'E}., \& {Hinderer}, T. 2008, \prd, 77, 021502

\bibitem[{{Fong} {et~al.}(2016){Fong}, {Metzger}, {Berger}, \&
  {{\"O}zel}}]{Fong16}
{Fong}, W., {Metzger}, B.~D., {Berger}, E., \& {{\"O}zel}, F. 2016, \apj, 831,
  141

\bibitem[{{Fong} {et~al.}(2012){Fong}, {Berger}, {Margutti}, {Zauderer},
  {Troja}, {Czekala}, {Chornock}, {Gehrels}, {Sakamoto}, {Fox}, \&
  {Podsiadlowski}}]{Fong12}
{Fong}, W., {Berger}, E., {Margutti}, R., {et~al.} 2012, \apj, 756, 189

\bibitem[{{Foucart}(2012)}]{Foucart12}
{Foucart}, F. 2012, \prd, 86, 124007

\bibitem[{{Friedman} \& {Schutz}(1978)}]{Friedman78}
{Friedman}, J.~L., \& {Schutz}, B.~F. 1978, \apj, 222, 281

\bibitem[{{Fryer} {et~al.}(2015){Fryer}, {Belczynski}, {Ramirez-Ruiz},
  {Rosswog}, {Shen}, \& {Steiner}}]{Fryer15}
{Fryer}, C.~L., {Belczynski}, K., {Ramirez-Ruiz}, E., {et~al.} 2015, \apj, 812,
  24

\bibitem[{{Gaertig} \& {Kokkotas}(2011)}]{Gaertig11}
{Gaertig}, E., \& {Kokkotas}, K.~D. 2011, \prd, 83, 064031

\bibitem[{{Gao} {et~al.}(2016){Gao}, {Zhang}, \& {L{\"u}}}]{Gao16}
{Gao}, H., {Zhang}, B., \& {L{\"u}}, H.-J. 2016, \prd, 93, 044065

\bibitem[{{Giacomazzo} \& {Perna}(2013)}]{Giacomazzo13}
{Giacomazzo}, B., \& {Perna}, R. 2013, \apjl, 771, L26

\bibitem[{{Giacomazzo} {et~al.}(2013){Giacomazzo}, {Perna}, {Rezzolla},
  {Troja}, \& {Lazzati}}]{Giacomazzo13a}
{Giacomazzo}, B., {Perna}, R., {Rezzolla}, L., {Troja}, E., \& {Lazzati}, D.
  2013, \apjl, 762, L18

\bibitem[{{Glendenning} \& {Moszkowski}(1991)}]{Glendenning91}
{Glendenning}, N.~K., \& {Moszkowski}, S.~A. 1991, Physical Review Letters, 67,
  2414

\bibitem[{{Hild} {et~al.}(2011){Hild}, {Abernathy}, {Acernese}, {Amaro-Seoane},
  {Andersson}, {Arun}, {Barone}, {Barr}, {Barsuglia}, {Beker}, {Beveridge},
  {Birindelli}, {Bose}, {Bosi}, {Braccini}, {Bradaschia}, {Bulik}, {Calloni},
  {Cella}, {Chassande Mottin}, {Chelkowski}, {Chincarini}, {Clark}, {Coccia},
  {Colacino}, {Colas}, {Cumming}, {Cunningham}, {Cuoco}, {Danilishin},
  {Danzmann}, {De Salvo}, {Dent}, {De Rosa}, {Di Fiore}, {Di Virgilio},
  {Doets}, {Fafone}, {Falferi}, {Flaminio}, {Franc}, {Frasconi}, {Freise},
  {Friedrich}, {Fulda}, {Gair}, {Gemme}, {Genin}, {Gennai}, {Giazotto},
  {Glampedakis}, {Gr{\"a}f}, {Granata}, {Grote}, {Guidi}, {Gurkovsky},
  {Hammond}, {Hannam}, {Harms}, {Heinert}, {Hendry}, {Heng}, {Hennes}, {Hough},
  {Husa}, {Huttner}, {Jones}, {Khalili}, {Kokeyama}, {Kokkotas}, {Krishnan},
  {Li}, {Lorenzini}, {L{\"u}ck}, {Majorana}, {Mandel}, {Mandic}, {Mantovani},
  {Martin}, {Michel}, {Minenkov}, {Morgado}, {Mosca}, {Mours},
  {M{\"u}ller-Ebhardt}, {Murray}, {Nawrodt}, {Nelson}, {Oshaughnessy}, {Ott},
  {Palomba}, {Paoli}, {Parguez}, {Pasqualetti}, {Passaquieti}, {Passuello},
  {Pinard}, {Plastino}, {Poggiani}, {Popolizio}, {Prato}, {Punturo}, {Puppo},
  {Rabeling}, {Rapagnani}, {Read}, {Regimbau}, {Rehbein}, {Reid}, {Ricci},
  {Richard}, {Rocchi}, {Rowan}, {R{\"u}diger}, {Santamar{\'{\i}}a}, {Sassolas},
  {Sathyaprakash}, {Schnabel}, {Schwarz}, {Seidel}, {Sintes}, {Somiya},
  {Speirits}, {Strain}, {Strigin}, {Sutton}, {Tarabrin}, {Th{\"u}ring}, {van
  den Brand}, {van Veggel}, {van den Broeck}, {Vecchio}, {Veitch}, {Vetrano},
  {Vicere}, {Vyatchanin}, {Willke}, {Woan}, \& {Yamamoto}}]{Hild11}
{Hild}, S., {Abernathy}, M., {Acernese}, F., {et~al.} 2011, Classical and
  Quantum Gravity, 28, 094013

\bibitem[{{Hinderer} {et~al.}(2010){Hinderer}, {Lackey}, {Lang}, \&
  {Read}}]{Hinderer10}
{Hinderer}, T., {Lackey}, B.~D., {Lang}, R.~N., \& {Read}, J.~S. 2010, \prd,
  81, 123016

\bibitem[{{Kalogera} {et~al.}(2004){Kalogera}, {Kim}, {Lorimer}, {Burgay},
  {D'Amico}, {Possenti}, {Manchester}, {Lyne}, {Joshi}, {McLaughlin}, {Kramer},
  {Sarkissian}, \& {Camilo}}]{Kalogera04}
{Kalogera}, V., {Kim}, C., {Lorimer}, D.~R., {et~al.} 2004, \apjl, 601, L179

\bibitem[{{Kastaun} {et~al.}(2016){Kastaun}, {Ciolfi}, \&
  {Giacomazzo}}]{Kastaun16}
{Kastaun}, W., {Ciolfi}, R., \& {Giacomazzo}, B. 2016, \prd, 94, 044060

\bibitem[{{Katz} {et~al.}(2009){Katz}, {Budnik}, \& {Waxman}}]{Katz09}
{Katz}, B., {Budnik}, R., \& {Waxman}, E. 2009, JCAP, 3, 020

\bibitem[{{Kawamura} {et~al.}(2016){Kawamura}, {Giacomazzo}, {Kastaun},
  {Ciolfi}, {Endrizzi}, {Baiotti}, \& {Perna}}]{Kawamura16}
{Kawamura}, T., {Giacomazzo}, B., {Kastaun}, W., {et~al.} 2016, \prd, 94,
  064012

\bibitem[{{Keane} {et~al.}(2012){Keane}, {Stappers}, {Kramer}, \&
  {Lyne}}]{Keane12}
{Keane}, E.~F., {Stappers}, B.~W., {Kramer}, M., \& {Lyne}, A.~G. 2012, \mnras,
  425, L71

\bibitem[{{Kim} {et~al.}(2005){Kim}, {Kalogera}, {Lorimer}, {Ihm}, \&
  {Belczynski}}]{Kim05}
{Kim}, C., {Kalogera}, V., {Lorimer}, D.~R., {Ihm}, M., \& {Belczynski}, K.
  2005, in Astronomical Society of the Pacific Conference Series, Vol. 328,
  Binary Radio Pulsars, ed. F.~A. {Rasio} \& I.~H. {Stairs}, 83

\bibitem[{{Kiziltan} {et~al.}(2013){Kiziltan}, {Kottas}, {De Yoreo}, \&
  {Thorsett}}]{Kiziltan13}
{Kiziltan}, B., {Kottas}, A., {De Yoreo}, M., \& {Thorsett}, S.~E. 2013, \apj,
  778, 66

\bibitem[{{Kochanek}(1992)}]{Kochanek92}
{Kochanek}, C.~S. 1992, \apj, 398, 234

\bibitem[{{Kopal}(1959)}]{Kopal59}
{Kopal}, Z. 1959, {Close binary systems}

\bibitem[{{Kotera} {et~al.}(2015){Kotera}, {Amato}, \& {Blasi}}]{Kotera15}
{Kotera}, K., {Amato}, E., \& {Blasi}, P. 2015, JCAP, 8, 026

\bibitem[{{Kotera} \& {Olinto}(2011)}]{Kotera11}
{Kotera}, K., \& {Olinto}, A.~V. 2011, \araa, 49, 119

\bibitem[{{Lai}(2001)}]{Lai01}
{Lai}, D. 2001, in American Institute of Physics Conference Series, Vol. 575,
  Astrophysical Sources for Ground-Based Gravitational Wave Detectors, ed.
  J.~M. {Centrella}, 246--257

\bibitem[{{Lasky} {et~al.}(2014){Lasky}, {Haskell}, {Ravi}, {Howell}, \&
  {Coward}}]{Lasky14}
{Lasky}, P.~D., {Haskell}, B., {Ravi}, V., {Howell}, E.~J., \& {Coward}, D.~M.
  2014, \prd, 89, 047302

\bibitem[{{Lattimer} \& {Prakash}(2001)}]{Lattimer01}
{Lattimer}, J.~M., \& {Prakash}, M. 2001, \apj, 550, 426

\bibitem[{{Law} {et~al.}(2009){Law}, {Kulkarni}, {Dekany}, {Ofek}, {Quimby},
  {Nugent}, {Surace}, {Grillmair}, {Bloom}, {Kasliwal}, {Bildsten}, {Brown},
  {Cenko}, {Ciardi}, {Croner}, {Djorgovski}, {van Eyken}, {Filippenko}, {Fox},
  {Gal-Yam}, {Hale}, {Hamam}, {Helou}, {Henning}, {Howell}, {Jacobsen},
  {Laher}, {Mattingly}, {McKenna}, {Pickles}, {Poznanski}, {Rahmer}, {Rau},
  {Rosing}, {Shara}, {Smith}, {Starr}, {Sullivan}, {Velur}, {Walters}, \&
  {Zolkower}}]{Law09}
{Law}, N.~M., {Kulkarni}, S.~R., {Dekany}, R.~G., {et~al.} 2009, \pasp, 121,
  1395

\bibitem[{{Lee} \& {Ramirez-Ruiz}(2007)}]{Lee07}
{Lee}, W.~H., \& {Ramirez-Ruiz}, E. 2007, New Journal of Physics, 9, 17

\bibitem[{{Letessier-Selvon} \& {Stanev}(2011)}]{Letessier11}
{Letessier-Selvon}, A., \& {Stanev}, T. 2011, Reviews of Modern Physics, 83,
  907

\bibitem[{{LIGO Scientific Collaboration} {et~al.}(2015){LIGO Scientific
  Collaboration}, {Aasi}, {Abbott}, {Abbott}, {Abbott}, {Abernathy}, {Ackley},
  {Adams}, {Adams}, {Addesso}, \& et~al.}]{LIGO}
{LIGO Scientific Collaboration}, {Aasi}, J., {Abbott}, B.~P., {et~al.} 2015,
  Classical and Quantum Gravity, 32, 074001

\bibitem[{{Lorimer} {et~al.}(2007){Lorimer}, {Bailes}, {McLaughlin},
  {Narkevic}, \& {Crawford}}]{Lorimer07}
{Lorimer}, D.~R., {Bailes}, M., {McLaughlin}, M.~A., {Narkevic}, D.~J., \&
  {Crawford}, F. 2007, Science, 318, 777

\bibitem[{{LSST Science Collaboration} {et~al.}(2009){LSST Science
  Collaboration}, {Abell}, {Allison}, {Anderson}, {Andrew}, {Angel}, {Armus},
  {Arnett}, {Asztalos}, {Axelrod}, \& et~al.}]{lsst}
{LSST Science Collaboration}, {Abell}, P.~A., {Allison}, J., {et~al.} 2009,
  ArXiv e-prints

\bibitem[{{L{\"u}} \& {Zhang}(2014)}]{Lu14}
{L{\"u}}, H.-J., \& {Zhang}, B. 2014, \apj, 785, 74

\bibitem[{{L{\"u}} {et~al.}(2015){L{\"u}}, {Zhang}, {Lei}, {Li}, \&
  {Lasky}}]{Lu15}
{L{\"u}}, H.-J., {Zhang}, B., {Lei}, W.-H., {Li}, Y., \& {Lasky}, P.~D. 2015,
  \apj, 805, 89

\bibitem[{{Martinez} {et~al.}(2015){Martinez}, {Stovall}, {Freire}, {Deneva},
  {Jenet}, {McLaughlin}, {Bagchi}, {Bates}, \& {Ridolfi}}]{Martinez15}
{Martinez}, J.~G., {Stovall}, K., {Freire}, P.~C.~C., {et~al.} 2015, \apj, 812,
  143

\bibitem[{{Metzger}(2016)}]{Metzger16}
{Metzger}, B.~D. 2016, ArXiv e-prints

\bibitem[{{Metzger} \& {Berger}(2012)}]{Metzger12}
{Metzger}, B.~D., \& {Berger}, E. 2012, \apj, 746, 48

\bibitem[{{Metzger} \& {Bower}(2014)}]{Metzger14b}
{Metzger}, B.~D., \& {Bower}, G.~C. 2014, \mnras, 437, 1821

\bibitem[{{Metzger} {et~al.}(2011){Metzger}, {Giannios}, {Thompson},
  {Bucciantini}, \& {Quataert}}]{Metzger11}
{Metzger}, B.~D., {Giannios}, D., {Thompson}, T.~A., {Bucciantini}, N., \&
  {Quataert}, E. 2011, \mnras, 413, 2031

\bibitem[{{Metzger} \& {Piro}(2014)}]{Metzger14}
{Metzger}, B.~D., \& {Piro}, A.~L. 2014, \mnras, 439, 3916

\bibitem[{{Miller} {et~al.}(2015){Miller}, {Barsotti}, {Vitale}, {Fritschel},
  {Evans}, \& {Sigg}}]{Miller15}
{Miller}, J., {Barsotti}, L., {Vitale}, S., {et~al.} 2015, \prd, 91, 062005

\bibitem[{{Murase} \& {Takami}(2009)}]{Murase09}
{Murase}, K., \& {Takami}, H. 2009, \apjl, 690, L14

\bibitem[{{Nakar}(2007)}]{Nakar07}
{Nakar}, E. 2007, \physrep, 442, 166

\bibitem[{{Nakar} \& {Piran}(2011)}]{Nakar11}
{Nakar}, E., \& {Piran}, T. 2011, \nat, 478, 82

\bibitem[{{Nicholl} {et~al.}(2017){Nicholl}, {Williams}, {Berger}, {Villar},
  {Alexander}, {Eftekhari}, \& {Metzger}}]{Nicholl17}
{Nicholl}, M., {Williams}, P.~K.~G., {Berger}, E., {et~al.} 2017, ArXiv
  e-prints

\bibitem[{{{\"O}zel} {et~al.}(2012){{\"O}zel}, {Psaltis}, {Narayan}, \& {Santos
  Villarreal}}]{Ozel12}
{{\"O}zel}, F., {Psaltis}, D., {Narayan}, R., \& {Santos Villarreal}, A. 2012,
  \apj, 757, 55

\bibitem[{{Passamonti} {et~al.}(2013){Passamonti}, {Gaertig}, {Kokkotas}, \&
  {Doneva}}]{Passamonti13}
{Passamonti}, A., {Gaertig}, E., {Kokkotas}, K.~D., \& {Doneva}, D. 2013, \prd,
  87, 084010

\bibitem[{{Passamonti} \& {Glampedakis}(2012)}]{Passamonti12}
{Passamonti}, A., \& {Glampedakis}, K. 2012, \mnras, 422, 3327

\bibitem[{{Piro} \& {Kollmeier}(2016)}]{Piro16b}
{Piro}, A.~L., \& {Kollmeier}, J.~A. 2016, \apj, 826, 97

\bibitem[{{Piro} \& {Kulkarni}(2013)}]{Piro13}
{Piro}, A.~L., \& {Kulkarni}, S.~R. 2013, \apjl, 762, L17

\bibitem[{{Popham} {et~al.}(1999){Popham}, {Woosley}, \& {Fryer}}]{Popham99}
{Popham}, R., {Woosley}, S.~E., \& {Fryer}, C. 1999, \apj, 518, 356

\bibitem[{{Price} \& {Rosswog}(2006)}]{Price06}
{Price}, D.~J., \& {Rosswog}, S. 2006, Science, 312, 719

\bibitem[{{Punturo} {et~al.}(2010){Punturo}, {Abernathy}, {Acernese}, {Allen},
  {Andersson}, {Arun}, {Barone}, {Barr}, {Barsuglia}, {Beker}, {Beveridge},
  {Birindelli}, {Bose}, {Bosi}, {Braccini}, {Bradaschia}, {Bulik}, {Calloni},
  {Cella}, {Chassande Mottin}, {Chelkowski}, {Chincarini}, {Clark}, {Coccia},
  {Colacino}, {Colas}, {Cumming}, {Cunningham}, {Cuoco}, {Danilishin},
  {Danzmann}, {De Luca}, {De Salvo}, {Dent}, {Derosa}, {Di Fiore}, {Di
  Virgilio}, {Doets}, {Fafone}, {Falferi}, {Flaminio}, {Franc}, {Frasconi},
  {Freise}, {Fulda}, {Gair}, {Gemme}, {Gennai}, {Giazotto}, {Glampedakis},
  {Granata}, {Grote}, {Guidi}, {Hammond}, {Hannam}, {Harms}, {Heinert},
  {Hendry}, {Heng}, {Hennes}, {Hild}, {Hough}, {Husa}, {Huttner}, {Jones},
  {Khalili}, {Kokeyama}, {Kokkotas}, {Krishnan}, {Lorenzini}, {L{\"u}ck},
  {Majorana}, {Mandel}, {Mandic}, {Martin}, {Michel}, {Minenkov}, {Morgado},
  {Mosca}, {Mours}, {M{\"u}ller-Ebhardt}, {Murray}, {Nawrodt}, {Nelson},
  {Oshaughnessy}, {Ott}, {Palomba}, {Paoli}, {Parguez}, {Pasqualetti},
  {Passaquieti}, {Passuello}, {Pinard}, {Poggiani}, {Popolizio}, {Prato},
  {Puppo}, {Rabeling}, {Rapagnani}, {Read}, {Regimbau}, {Rehbein}, {Reid},
  {Rezzolla}, {Ricci}, {Richard}, {Rocchi}, {Rowan}, {R{\"u}diger}, {Sassolas},
  {Sathyaprakash}, {Schnabel}, {Schwarz}, {Seidel}, {Sintes}, {Somiya},
  {Speirits}, {Strain}, {Strigin}, {Sutton}, {Tarabrin}, {van den Brand}, {van
  Leewen}, {van Veggel}, {van den Broeck}, {Vecchio}, {Veitch}, {Vetrano},
  {Vicere}, {Vyatchanin}, {Willke}, {Woan}, {Wolfango}, \&
  {Yamamoto}}]{Punturo10}
{Punturo}, M., {Abernathy}, M., {Acernese}, F., {et~al.} 2010, Classical and
  Quantum Gravity, 27, 084007

\bibitem[{{Qian} \& {Wasserburg}(2007)}]{Qian07}
{Qian}, Y.-Z., \& {Wasserburg}, G.~J. 2007, \physrep, 442, 237

\bibitem[{{Rane} {et~al.}(2016){Rane}, {Lorimer}, {Bates}, {McMann},
  {McLaughlin}, \& {Rajwade}}]{Rane16}
{Rane}, A., {Lorimer}, D.~R., {Bates}, S.~D., {et~al.} 2016, \mnras, 455, 2207

\bibitem[{{Ravi} {et~al.}(2015){Ravi}, {Shannon}, \& {Jameson}}]{Ravi15}
{Ravi}, V., {Shannon}, R.~M., \& {Jameson}, A. 2015, \apjl, 799, L5

\bibitem[{{Read} {et~al.}(2009){Read}, {Markakis}, {Shibata}, {Ury{\= u}},
  {Creighton}, \& {Friedman}}]{Read09}
{Read}, J.~S., {Markakis}, C., {Shibata}, M., {et~al.} 2009, \prd, 79, 124033

\bibitem[{{Rezzolla} {et~al.}(2011){Rezzolla}, {Giacomazzo}, {Baiotti},
  {Granot}, {Kouveliotou}, \& {Aloy}}]{Rezzolla11}
{Rezzolla}, L., {Giacomazzo}, B., {Baiotti}, L., {et~al.} 2011, \apjl, 732, L6

\bibitem[{{Rosswog}(2015)}]{Rosswog15}
{Rosswog}, S. 2015, International Journal of Modern Physics D, 24, 1530012

\bibitem[{{Rowlinson} {et~al.}(2013){Rowlinson}, {O'Brien}, {Metzger},
  {Tanvir}, \& {Levan}}]{Rowlinson13}
{Rowlinson}, A., {O'Brien}, P.~T., {Metzger}, B.~D., {Tanvir}, N.~R., \&
  {Levan}, A.~J. 2013, \mnras, 430, 1061

\bibitem[{{Rowlinson} {et~al.}(2010){Rowlinson}, {O'Brien}, {Tanvir}, {Zhang},
  {Evans}, {Lyons}, {Levan}, {Willingale}, {Page}, {Onal}, {Burrows},
  {Beardmore}, {Ukwatta}, {Berger}, {Hjorth}, {Fruchter}, {Tunnicliffe}, {Fox},
  \& {Cucchiara}}]{Rowlinson10}
{Rowlinson}, A., {O'Brien}, P.~T., {Tanvir}, N.~R., {et~al.} 2010, \mnras, 409,
  531

\bibitem[{{Ruiz} {et~al.}(2016){Ruiz}, {Lang}, {Paschalidis}, \&
  {Shapiro}}]{Ruiz16}
{Ruiz}, M., {Lang}, R.~N., {Paschalidis}, V., \& {Shapiro}, S.~L. 2016, \apjl,
  824, L6

\bibitem[{{Scholz} {et~al.}(2016){Scholz}, {Spitler}, {Hessels}, {Chatterjee},
  {Cordes}, {Kaspi}, {Wharton}, {Bassa}, {Bogdanov}, {Camilo}, {Crawford},
  {Deneva}, {van Leeuwen}, {Lynch}, {Madsen}, {McLaughlin}, {Mickaliger},
  {Parent}, {Patel}, {Ransom}, {Seymour}, {Stairs}, {Stappers}, \&
  {Tendulkar}}]{Scholz16}
{Scholz}, P., {Spitler}, L.~G., {Hessels}, J.~W.~T., {et~al.} 2016, ArXiv
  e-prints

\bibitem[{{Shapiro}(2000)}]{Shapiro00}
{Shapiro}, S.~L. 2000, \apj, 544, 397

\bibitem[{{Shapiro} \& {Teukolsky}(1983)}]{Shapiro83}
{Shapiro}, S.~L., \& {Teukolsky}, S.~A. 1983, {Black holes, white dwarfs, and
  neutron stars: The physics of compact objects}

\bibitem[{{Shappee} {et~al.}(2014){Shappee}, {Prieto}, {Grupe}, {Kochanek},
  {Stanek}, {De Rosa}, {Mathur}, {Zu}, {Peterson}, {Pogge}, {Komossa}, {Im},
  {Jencson}, {Holoien}, {Basu}, {Beacom}, {Szczygie{\l}}, {Brimacombe},
  {Adams}, {Campillay}, {Choi}, {Contreras}, {Dietrich}, {Dubberley},
  {Elphick}, {Foale}, {Giustini}, {Gonzalez}, {Hawkins}, {Howell}, {Hsiao},
  {Koss}, {Leighly}, {Morrell}, {Mudd}, {Mullins}, {Nugent}, {Parrent},
  {Phillips}, {Pojmanski}, {Rosing}, {Ross}, {Sand}, {Terndrup}, {Valenti},
  {Walker}, \& {Yoon}}]{Shappee14}
{Shappee}, B.~J., {Prieto}, J.~L., {Grupe}, D., {et~al.} 2014, \apj, 788, 48

\bibitem[{{Shibata} {et~al.}(2000){Shibata}, {Baumgarte}, \&
  {Shapiro}}]{Shibata00}
{Shibata}, M., {Baumgarte}, T.~W., \& {Shapiro}, S.~L. 2000, \apj, 542, 453

\bibitem[{{Shibata} \& {Taniguchi}(2011)}]{Shibata11}
{Shibata}, M., \& {Taniguchi}, K. 2011, Living Reviews in Relativity, 14, 6

\bibitem[{{Siegel} \& {Ciolfi}(2016{\natexlab{a}})}]{Siegel16a}
{Siegel}, D.~M., \& {Ciolfi}, R. 2016{\natexlab{a}}, \apj, 819, 14

\bibitem[{{Siegel} \& {Ciolfi}(2016{\natexlab{b}})}]{Siegel16b}
---. 2016{\natexlab{b}}, \apj, 819, 15

\bibitem[{{Sneden} {et~al.}(2008){Sneden}, {Cowan}, \& {Gallino}}]{Sneden08}
{Sneden}, C., {Cowan}, J.~J., \& {Gallino}, R. 2008, \araa, 46, 241

\bibitem[{{Spitler} {et~al.}(2014){Spitler}, {Cordes}, {Hessels}, {Lorimer},
  {McLaughlin}, {Chatterjee}, {Crawford}, {Deneva}, {Kaspi}, {Wharton},
  {Allen}, {Bogdanov}, {Brazier}, {Camilo}, {Freire}, {Jenet},
  {Karako-Argaman}, {Knispel}, {Lazarus}, {Lee}, {van Leeuwen}, {Lynch},
  {Ransom}, {Scholz}, {Siemens}, {Stairs}, {Stovall}, {Swiggum},
  {Venkataraman}, {Zhu}, {Aulbert}, \& {Fehrmann}}]{Spitler14}
{Spitler}, L.~G., {Cordes}, J.~M., {Hessels}, J.~W.~T., {et~al.} 2014, \apj,
  790, 101

\bibitem[{{Spitler} {et~al.}(2016){Spitler}, {Scholz}, {Hessels}, {Bogdanov},
  {Brazier}, {Camilo}, {Chatterjee}, {Cordes}, {Crawford}, {Deneva}, {Ferdman},
  {Freire}, {Kaspi}, {Lazarus}, {Lynch}, {Madsen}, {McLaughlin}, {Patel},
  {Ransom}, {Seymour}, {Stairs}, {Stappers}, {van Leeuwen}, \&
  {Zhu}}]{Spitler16}
{Spitler}, L.~G., {Scholz}, P., {Hessels}, J.~W.~T., {et~al.} 2016, \nat, 531,
  202

\bibitem[{{Tanvir} {et~al.}(2013){Tanvir}, {Levan}, {Fruchter}, {Hjorth},
  {Hounsell}, {Wiersema}, \& {Tunnicliffe}}]{Tanvir13}
{Tanvir}, N.~R., {Levan}, A.~J., {Fruchter}, A.~S., {et~al.} 2013, \nat, 500,
  547

\bibitem[{{The IceCube Collaboration}(2011)}]{IceCube}
{The IceCube Collaboration}. 2011, ArXiv e-prints

\bibitem[{{Thielemann} {et~al.}(2011){Thielemann}, {Arcones}, {K{\"a}ppeli},
  {Liebend{\"o}rfer}, {Rauscher}, {Winteler}, {Fr{\"o}hlich}, {Dillmann},
  {Fischer}, {Martinez-Pinedo}, {Langanke}, {Farouqi}, {Kratz}, {Panov}, \&
  {Korneev}}]{Thielemann11}
{Thielemann}, F.-K., {Arcones}, A., {K{\"a}ppeli}, R., {et~al.} 2011, Progress
  in Particle and Nuclear Physics, 66, 346

\bibitem[{{Thornton} {et~al.}(2013){Thornton}, {Stappers}, {Bailes},
  {Barsdell}, {Bates}, {Bhat}, {Burgay}, {Burke-Spolaor}, {Champion}, {Coster},
  {D'Amico}, {Jameson}, {Johnston}, {Keith}, {Kramer}, {Levin}, {Milia}, {Ng},
  {Possenti}, \& {van Straten}}]{Thornton13}
{Thornton}, D., {Stappers}, B., {Bailes}, M., {et~al.} 2013, Science, 341, 53

\bibitem[{{Timmes} {et~al.}(1996){Timmes}, {Woosley}, \& {Weaver}}]{Timmes96}
{Timmes}, F.~X., {Woosley}, S.~E., \& {Weaver}, T.~A. 1996, \apj, 457, 834

\bibitem[{{Usov}(1992)}]{Usov92}
{Usov}, V.~V. 1992, \nat, 357, 472

\bibitem[{{Waxman}(1995)}]{Waxman95}
{Waxman}, E. 1995, Physical Review Letters, 75, 386

\bibitem[{{Zhang} \& {M{\'e}sz{\'a}ros}(2001)}]{Zhang01}
{Zhang}, B., \& {M{\'e}sz{\'a}ros}, P. 2001, \apjl, 552, L35

\bibitem[{{Zrake} \& {MacFadyen}(2013)}]{Zrake13}
{Zrake}, J., \& {MacFadyen}, A.~I. 2013, \apjl, 769, L29

\end{thebibliography}

\end{document}